\documentclass[preprint,showkeys,preprintnumbers,amsmath,amssymb]{revtex4}

\usepackage{graphicx}
\usepackage{dcolumn}
\usepackage{bm}

\begin{document}

\title{ Towards a Field Theoretical Stochastic
Model \\	[2mm] for Description of Tumour Growth}

\author{Leonardo Mondaini}
\email{mondaini@unirio.br}
\affiliation{Department of Physics, Federal University of the State of Rio de Janeiro - UNIRIO, \\Rio de Janeiro, Brazil}

\begin{abstract}
We develop a field theory-inspired stochastic model
for description of tumour growth based on an
analogy with an SI epidemic model, where the
susceptible individuals (S) would represent the
healthy cells and the infected ones (I), the cancer
cells. From this model, we obtain a curve
describing the tumour volume as a function of
time, which can be compared to available
experimental data.
\end{abstract}


\maketitle

\section{Introduction}

In a recent letter \cite{mondaini}, we have shown how the standard field theoretical language based on creation and annihilation operators (building blocks of the second quantization method \cite{greiner}) 
may be used for a straightforward derivation of closed master equations \cite{gardiner} describing the population dynamics of multivariate stochastic epidemic models. This was mainly motivated by the observation that, as remarked in \cite{dodd}, for the kinds of model studied in population biology and epidemiology, a field theoretical description is notationally neater and more manageable than standard methods, in often replacing sets of equations with single equations with the same content. Indeed, a single hamiltonian function sums up the dynamics compactly, even when births and deaths allow the population size to change, and may be easily written down from a verbal description of the transitions presented in these models.

In the present work we employ the very same methodology established in \cite{mondaini} to develop a field-theory inspired stochastic model for description of tumour growth based on an analogy with an SI epidemic model \cite{allen}, where the susceptible individuals (S) would represent the healthy cells and the infected ones (I), the cancer cells. From this model, we were able to obtain a curve describing the time evolution of the tumour volume, which is then compared to available
experimental data for liver cancer. Our main motivation comes from the pioneering ideas about cancer as a phase transition presented in \cite{jack}, more specifically, from the following observations: (1) progression of cancer must also involve a population-level shift due to a competition between two co-existing phenotypes: {\it normal} and {\it cancerous}, and (2) given enough time and resources, cancer cells will usually outcompete healthy cells in the organ or tissue where they coexist, in the competiton for space and resources.

The rest of this work is organized as follows. In Section 2, we introduce the basic aspects of our model which allow us to obtain differential equations describing the time evolution of the mean number of individuals in the interacting populations we are dealing with, i.e., normal (healthy) and cancer cells. The analytical solutions for these differential equations, besides the specific curve describing the time evolution of tumour volume and its comparison to experimental data are presented in Section 3. Finally, in Section 4 we present our concluding remarks. 

\section{Building the model}
We will start by considering interacting populations, whose total sizes are allowed to change, composed of two types of individuals: 
the normal (healthy) cells and the tumour (cancer) cells. Let us introduce $\mathcal{N}(t)$ and $\mathcal{C}(t)$ as random variables which represent, respectively, the number of normal and cancer cells at a given time instant $t$.

We will then consider a bivariate process $\{(\mathcal{N}(t);\,\mathcal{C}(t))\}_{t=0}^\infty$ with a joint probability function given by

\begin{equation}
p_{(n,m)}(t)={\rm{Prob}}\{\mathcal{N}(t)=n;\,\mathcal{C}(t)=m\}. \label{eq1}
\end{equation}

Our aim is to compute time-dependent expectation values of the observables $\mathcal{N}(t)$ and $\mathcal{C}(t)$, which may be defined in terms of the configuration probability according to
\begin{eqnarray}
\left<\mathcal{N}(t)\right>=\sum_{n,m}n\,p_{(n,m)}(t);\nonumber\\
 \left<\mathcal{C}(t)\right>=\sum_{n,m}m\,p_{(n,m)}(t). \label{eq2}
\end{eqnarray}
Let us represent the probabilistic state of the system by the vector
\begin{equation}
\left|\mu,\nu\right>=\sum_{n,m} p_{(n,m)}(t)\left|n, m\right>, \label{eq3}
\end{equation}
with the normalization condition $\sum_{n,m} p_{(n,m)}(t)=1$.

As an example, the vector $\frac14\left(\left|1,1\right>+\left|2,1\right>+\left|1,2\right>+\left|2,2\right>\right)$ represents the probability 
distribution where there are 1 or 2 healthy/cancer cells present, each one with probability 1/4, i.e. $p_{(1,1)}=p_{(2,1)}=p_{(1,2)}=p_{(2,2)}=\frac14$.

Since the configurations are given entirely in terms of occupation numbers ($n, m$), which calls for a representation in terms of second-quantized 
bosonic operators \cite{cardy}, we will introduce {\it creation} and {\it annihilation} operators for the normal cells, respectively, $h^\dagger$ and $h$, and for the cancer cells, namely, $c^\dagger$ and $c$.
These operators must satisfy the following commutation relations
\begin{eqnarray}
\left[h,\,h^\dagger\right] = \left[c,\,c^\dagger\right] = 1;\nonumber \\
\left[h,\,c\right] = \left[h,\,c^\dagger\right] =\left[c,\,h^\dagger\right] =\left[h^\dagger,\,c^\dagger\right]  = 0. \label{eq4}
\end{eqnarray}
As usual in the second quantization framework, we say that $h^\dagger$ and $c^\dagger$ ``create" , respectively, normal and cancer cells when applied over the reference (vacuum) state $\left|0,0\right>$. This allows us to build our space from basis vectors of the form $\left|n,m\right>=\left(h^\dagger\right)^n\left(c^\dagger\right)^m\left|0, 0\right>$.

This vacuum state has the following properties: $h\left|0,0\right>=c\left|0,0\right>=0$ (from which ``annihilation" operators) and $\left<0,0|0,0\right>=1$ (inner product).

Following the above definitions, we also have
\begin{eqnarray}
 h^\dagger\,\left|n, m\right> = \left|n+1, m\right>;\,\,\,\,\,\,c^\dagger\,\left|n, m\right> = \left|n, m+1\right>;\nonumber\\
h\,\left|n, m\right> = n\left|n-1, m\right>;\,\,\,\,\,\,c\,\left|n, m\right> = m\left|n, m-1\right>. \label{eq5}
\end{eqnarray}
At this point it is worth to note that $h^\dagger h\,\left|n, m\right> = n\left|n, m\right>$ and $c^\dagger c\,\left|n, m\right> = m\left|n, m\right>$. Thus, the operators $n=h^\dagger h$ and $m=c^\dagger c$ just count the number of cells in a definite state. This is the main reason why they are usually called {\it number operators}.
The vector state of our system may be then rewritten in terms of creation and annihilation operators as
\begin{eqnarray}
\left|\mu,\nu\right>=\sum_{n,m} p_{(n,m)}(t) \left(h^\dagger\right)^n\left(c^\dagger\right)^m\left|0, 0\right>. \label{eq6}
\end{eqnarray}

The time evolution of our system will then be generated by a linear operator $\mathcal{H}$ (called {\it hamiltonian}) which may be constructed directly from the transition rates present in our model according to Table \ref{table1} (cf. \cite{dodd}, Table 1). Note that, upon summing up the terms presented in Table \ref{table1}, we may write our hamiltonian as
\begin{eqnarray}
	\mathcal{H} =+\left(b_h+d_h+\lambda\right) n + \left(b_c+d_c\right) m \nonumber \\
       -\left(b_h h^\dagger n + b_c c^\dagger m + d_h h + d_c c + \lambda c^\dagger h\right). \label{eq7}
\end{eqnarray}

\begin{table}
\centering\caption{Transition rates presented in our model and corresponding terms in the hamiltonian $\mathcal{H}$.}
{\begin{tabular}{l l l}
\hline\hline
{Transition} & {Description} & {Contribution to $\mathcal{H}$}\\
\hline
$\mathcal{N} \xrightarrow{b_h} \mathcal{N}+\mathcal{N}$ & birth of normal cell (rate $b_h$)&$b_h(h^\dagger h-h^\dagger h^\dagger h)=b_h(n-h^\dagger n)$\\
$\mathcal{C} \xrightarrow{b_c} \mathcal{C}+\mathcal{C}$ & birth of cancer cell (rate $b_c$)&$b_c(c^\dagger c-c^\dagger c^\dagger c)=b_c(m-c^\dagger m)$\\
$\mathcal{N} \xrightarrow{d_h} \varnothing $ & death of normal cell (rate $d_h$)&$d_h(h^\dagger h-h)=d_h(n-h)$\\
$\mathcal{C} \xrightarrow{d_c} \varnothing $ & death of cancer cell (rate $d_c$)&$d_c(c^\dagger c-c)=d_c(m-c)$\\
$\mathcal{N} \xrightarrow{\lambda} \mathcal{C}$ & change normal $\rightarrow$ cancer (rate $\lambda$)&$\lambda(h^\dagger h-c^\dagger h)=\lambda(n-c^\dagger h)$\\
\hline
\end{tabular}}
\label{table1}
\end{table}

The notational advantage of this field theoretical description is made clear at this point if we observe that, from the above definitions, the equation which 
represents the flux of probability between states at rates defined by our model (the so-called {\it master equation} or 
{\it forward Kolmogorov differential equation} \cite{allen}) takes the very compact form of an imaginary-time Schr\"odinger-type linear equation, namely
\begin{eqnarray}
\frac{d}{dt}\left|\mu,\nu\right> = -\mathcal{H}\left|\mu,\nu\right>. \label{eq8}
\end{eqnarray}
We can get, after some algebra, a more common representation for the master equation by substituting the expressions for the hamiltonian  
(\ref{eq7}) and the vector state (\ref{eq3}) into Equation (\ref{eq8})
\begin{eqnarray}
\frac{d}{dt}p_{(n,m)}=- [(b_h+d_h+\lambda) n + (b_c+d_c) m]\,p_{(n,m)}\nonumber\\
+ b_h(n-1)\,p_{(n-1,m)}+b_c(m-1)\,p_{(n,m-1)}\nonumber\\
+ d_h(n+1) \,p_{(n+1,m)}+d_c(m+1) \,p_{(n,m+1)}\nonumber\\
+ \lambda (n+1)\,p_{(n+1,m-1)}. \label{eq9}
\end{eqnarray}

In order to compute the time-dependent expectation values of the observables $\mathcal{N}(t)$ and $\mathcal{C}(t)$ through the above master equation, 
we will follow the well-established methodology presented in \cite{allen} and introduce the following {\it moment generating function (mgf)}
\begin{eqnarray}
M(\phi,\theta;t)=\left<e^{\phi \mathcal{N}(t)}e^{\theta \mathcal{C}(t)}\right> = \sum_{n,m} p_{(n,m)}e^{n\phi+m\theta}.\label{eq10}
\end{eqnarray}
Note that from the above equation we have
\begin{eqnarray}
\left[\frac{\partial M}{\partial\phi}\right]_{\phi,\theta=0} =  \sum_{n,m} n\, p_{(n,m)} = \left<\mathcal{N}(t)\right>;\nonumber \\ \left[\frac{\partial M}{\partial\theta}\right]_{\phi,\theta=0} =  \sum_{n,m} m\, p_{(n,m)} = \left<\mathcal{C}(t)\right>\label{eq11}
\end{eqnarray}
and, in general
\begin{eqnarray}
\left[\frac{\partial^k M}{\partial\theta^k}\right]_{\phi,\theta=0} = \left<\mathcal{C}^k(t)\right>;\,\,\,\,\,\,\,\,\,\,\,\,\,\,\,\,\,\left[\frac{\partial^k M}{\partial\phi^k}\right]_{\phi,\theta=0} = \left<\mathcal{N}^k(t)\right>.\label{eq12}
\end{eqnarray}

After multiplying Equation (\ref{eq9}) by $\exp(n\phi+m\theta)$ and summing on ($n, m$), we are led, after some algebra, to
\begin{eqnarray}
\frac{\partial M}{\partial t} = \sum_{n,m=0} \frac{d\,p_{(n,m)}}{dt}\,e^{n\phi + m\theta}\nonumber\\
= + \left[b_h\left(e^\phi - 1\right) + d_h\left(e^{-\phi} - 1\right)+\lambda\left(e^{-\phi+\theta} - 1\right)\right]\frac{\partial M}{\partial\phi}\nonumber\\
+  \left[b_c\left(e^\theta - 1\right) + d_c\left(e^{-\theta} - 1\right)\right]\frac{\partial M}{\partial\theta}.\label{eq13}
\end{eqnarray}
Finally, by  differentiating the above equation with respect to $\phi$ and evaluating the result at $\phi=\theta=0$ we get the following differential equation for $\left<\mathcal{N}(t)\right>$
\begin{eqnarray}
\left[\frac{\partial^2 M}{\partial t \,\partial \phi}\right]_{\phi,\theta = 0} = \frac{d}{dt}\left<\mathcal{N}(t)\right> = (b_h-d_h-\lambda) \left<\mathcal{N}(t)\right>. \label{eq14}
\end{eqnarray}
On the other hand, if we differentiate Equation (\ref{eq13}) with respect to $\theta$  and evaluate at $\phi=\theta=0$ we get the following differential equation for $\left<\mathcal{C}(t)\right>$
\begin{eqnarray}
\left[\frac{\partial^2 M}{\partial t\, \partial \theta}\right]_{\phi,\theta = 0} = \frac{d}{dt}\left<\mathcal{C}(t)\right> = \lambda \left<\mathcal{N}(t)\right> + (b_c-d_c) \left<\mathcal{C}(t)\right>.\label{eq15}
\end{eqnarray}

\section{Comparison to liver cancer data. Fitting the parameters in the analytical solution for $ \left<\mathcal{C}(t)\right>$}

By defining $\left<\mathcal{N}(0)\right>\equiv N_0$, $\left<\mathcal{C}(0)\right>\equiv C_0$, $\beta_h \equiv b_h - d_h$ and $\beta_c \equiv b_c - d_c$ 
we obtain the following analytical solutions for Equations (\ref{eq14}) and (\ref{eq15})
\begin{eqnarray}
\left<\mathcal{N}(t)\right>=N_0\,e^{(\beta_h-\lambda)t} \label{eqSolutionN}
\end{eqnarray}
and
\begin{eqnarray}
\left<\mathcal{C}(t)\right>=\frac{\lambda N_0}{\beta_h-\lambda-\beta_c}\left[e^{(\beta_h-\lambda)t}-e^{\beta_c t}\right] + C_0 e^{\beta_c t}. \label{eqSolutionC}
\end{eqnarray}
Finally, if we consider that the number of normal cells is approximately constant $(\beta_h\approx 0)$ and that the volume of a cancer cell ($v$) is approximately the same of a normal cell, we may write the following expression for the time evolution of the tumour volume
\begin{eqnarray}
V_c(t)=\frac{\lambda (V_h)_0}{-\lambda-\beta_c}\left[e^{-\lambda t}-e^{\beta_c t}\right] + (V_c)_0 e^{\beta_c t}, \label{eqSolutionVc}
\end{eqnarray}
where $V_c(t)\equiv v\left<\mathcal{C}(t)\right>$, $(V_h)_0\equiv v N_0$ is the initial volume of the normal tissue, and  $(V_c)_0\equiv v C_0$ is the initial tumour volume.

\begin{figure} [h]
\begin{center}
\hfil\scalebox{0.4}{\includegraphics{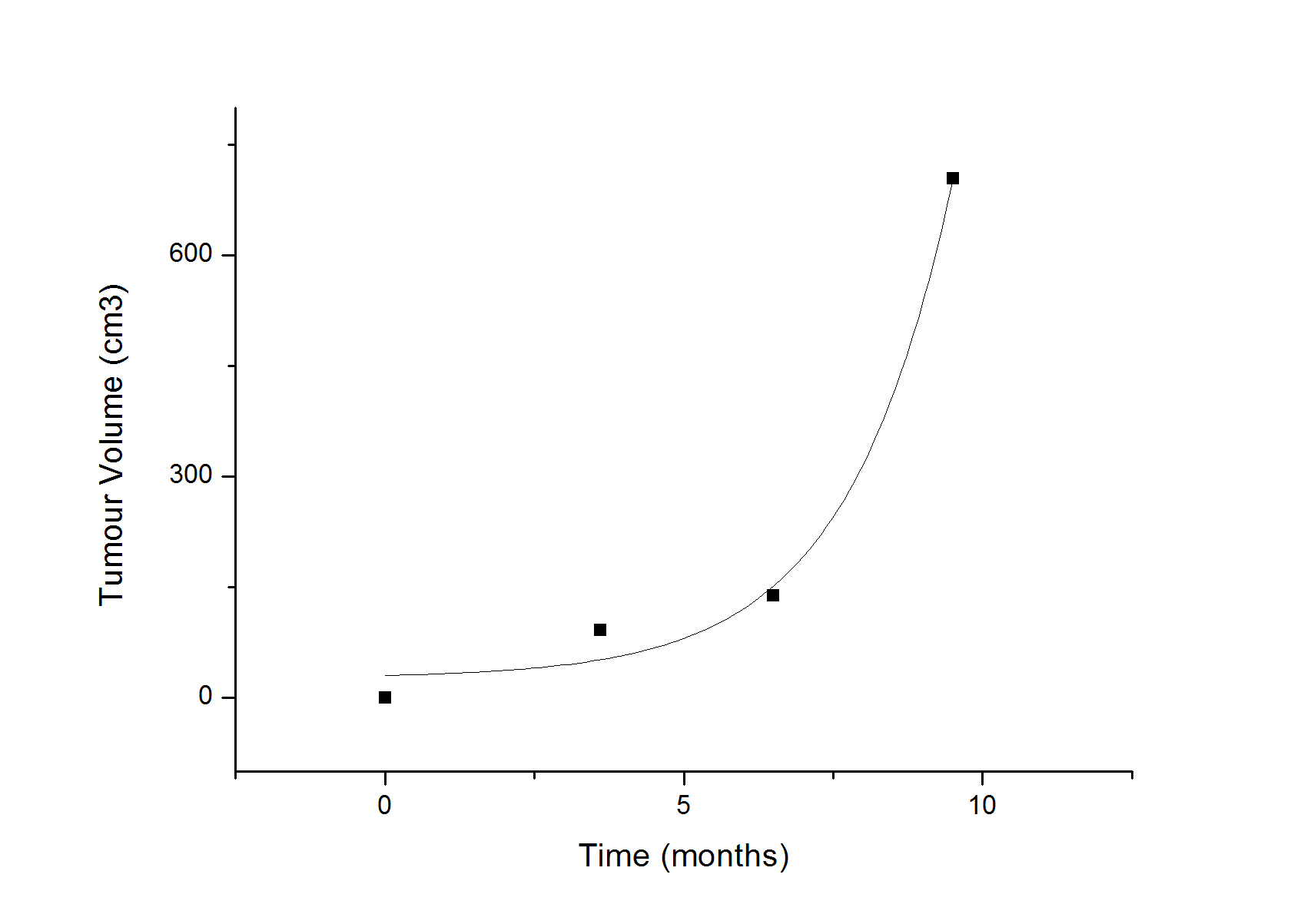}}\hfil
\caption{The solid line represents Equation (\ref{eqSolutionVc}) for an initial total volume of the patient liver $(V_T)_0 = (V_h)_0 + (V_c)_0 = 2153.00\, {\rm{cm}}^3$; $\lambda = -7.07\times 10^{-3}\,{\rm{months}}^{-1};  \,\beta_c = 5.67\times 10^{-1}\, {\rm{months}}^{-1}$ and $(V_c)_0 = 29.87\, {\rm{cm}}^3$. This curve was obtained by fitting the experimental data available in \cite{data} (square points in the figure), using the software OriginPro 8 \cite{origin}.}
\label{Figure1}
\end{center}
\end{figure}

The above expression is compared to experimental data for liver cancer (average tumour volume) in Figure \ref{Figure1}.  The corresponding data have been obtained through the analysis of computed tomography (CT) scans for a set of 34 patients available in \cite{data}.  

\section{Concluding Remarks}

We would like to finish this work by making a few comments about Equation (\ref{eqSolutionVc}). As far as we know, this is the first time that an expression describing the
time evolution of the volume of tumours \footnote{A comprehensive review of the most common mathematical models for description of tumour growth may be found in \cite{benzekry}.} is obtained from basic assumptions about cancer as a phase transition and, most certainly, this is the very first model for 
description of tumour growth built by using standard field theoretical language commonly found in models describing fundamental interactions of elementary particles. 
In a future work we are going to present a qualitative analysis of the behaviour of solutions for the system of first order linear differential equations composed by 
(\ref{eq14}) and (\ref{eq15}) in the phase plane, which we believe will shed more light on how our model is indeed connected to the idea of cancer as a phase transition.

Last but not least, other possible extensions for the present model should consider the inclusion of other kinds of dependence in the birth/death rates of cells, as temperature 
and/or concentration gradients of toxic carcinogens, for example.

\begin{acknowledgments}
The author would like to thank Prof. Jack A. Tuszynski for the valuable discussions and the very stimulating scientific environment shared at
University of Alberta's Li Ka Shing Applied Virology Institute, where the main results of this work were obtained.
\end{acknowledgments}

\end{document}